\documentclass[aps, pre, reprint, superscriptaddress, amsmath, amssymb, floatfix]{revtex4-2}

\usepackage{graphicx}
\usepackage{dcolumn}
\usepackage{bm}
\usepackage{hyperref}
\usepackage{cleveref}
\usepackage{xcolor}
\usepackage[protrusion=true,expansion=true]{microtype}
\DisableLigatures{encoding = *, family = *} 
\usepackage{multirow}
\usepackage{makecell}

\def\min{\textrm{min}}

\def\T{\mathcal{T}}

\def\homogeneous networks{\text{homogeneous networks}}

\def\BS{\text{BS}}
\def\BA{\text{BA}}
\def\E{\mathcal{E}}         
\def\PM{\text{PM}}          
\def\Baxter{\text{Baxter}}          
\def\Baxter{\text{Baxter}}  
\def\$\langle \sigma \rangle${{\langle \sigma \rangle}}          
\def\AF{\text{AF}}          
\def\kbar{\langle k \rangle}     
\def\kkbar{\langle k^2 \rangle}  

\begin{document}

\preprint{APS/123-QED}

\title{Ashkin-Teller model with antiferromagnetic 
four-spin interactions: \\ Interference effect between two conflicting issues}
\author{Cook Hyun Kim}
\affiliation{Center for Complex Systems, KI of Grid Modernization, Korea Institute of Energy Technology, Naju, Jeonnam 58330, Korea}
\author{Hoyun Choi}
\affiliation{Center for Complex Systems, KI of Grid Modernization, Korea Institute of Energy Technology, Naju, Jeonnam 58330, Korea}
\author{Joonsung Jung}
\affiliation{Center for Complex Systems, KI of Grid Modernization, Korea Institute of Energy Technology, Naju, Jeonnam 58330, Korea}
\author{B. Kahng}
\email{bkahng@kentech.ac.kr}
\affiliation{Center for Complex Systems, KI of Grid Modernization, Korea Institute of Energy Technology, Naju, Jeonnam 58330, Korea}
\date{\today}

\begin{abstract}
Spin systems have emerged as powerful tools for understanding collective phenomena in complex systems. In this work, we investigate the Ashkin--Teller (AT) model on random scale-free networks using mean-field theory, which extends the traditional Ising framework by coupling two spin systems via both pairwise and four-spin interactions. We focus on the previously unexplored antiferromagnetic regime of four-spin coupling, in which strong ordering in one layer actively suppresses the formation of order in the other layer. This mechanism captures, for example, scenarios in social or political systems where a dominant viewpoint on one issue (e.g., economic development) can inhibit consensus on another (e.g., environmental conservation). Our analysis reveals a rich phase diagram with four distinct phases---paramagnetic, Baxter, $\langle \sigma \rangle$, and antiferromagnetic---and diverse types of phase transitions. Notably, we find that the upper critical degree exponent extends to $\lambda_{c2} \approx 9.237$, far exceeding the conventional value of $\lambda = 5$ observed in ferromagnetic systems. This dramatic shift underscores the enhanced robustness of hub-mediated spin correlations under competitive coupling, leading to asymmetric order parameters between layers and novel phase transition phenomena. These findings offer fundamental insights into systems with competing order parameters and have direct implications for multilayer biological networks, social media ecosystems, and political debates characterized by competing priorities.
\end{abstract}

\maketitle

\section{Introduction}\label{sec:intro}

The Ising model's binary spin variables and their interactions have become a fundamental paradigm for understanding collective behavior in complex systems~\cite{stanley1971phase, chaikin1995principles}. While originally developed for magnetic materials, this framework now illuminates emergent phenomena across diverse fields, from financial markets~\cite{bouchaud2013crises, sornette2014physics} to neural networks~\cite{carreira2005contrastive, schneidman2006weak}, with particularly profound implications for social systems where individual interactions drive collective patterns~\cite{castellano2009statistical, stauffer2013biased, schweitzer2018sociophysics, sen2014sociophysics}.

Within social network analysis, the traditional Ising model~\cite{brush1967history} has proven especially useful by naturally mapping opposing viewpoints onto spin states~\cite{redner2019reality, clifford1973model, holley1975ergodic, cox1983occupation}. This framework gains additional relevance when implemented on random scale-free (SF) networks~\cite{dorogovtsev2002ising, igloi2002first, leone2002ferromagnetic, bianconi2002mean, herrero2004ising, lee2009critical}, whose heterogeneous connectivity patterns---characterized by highly connected hub nodes---mirror the structure of many real social systems. The Ashkin-Teller (AT) model extends this paradigm by coupling two Ising systems through both pairwise ($J_2$) and four-spin ($J_4$) interactions~\cite{ashkin1943statistics, kadanoff1971some, fan1972symmetry, wegner1972duality, ditzian1980phase, kohmoto1981hamiltonian, jang2015ashkin, kim2021link}:
\begin{align}
\mathcal{H} 
= - \sum_{\langle i,j \rangle} \Big[J_2\big(s_i s_j + \sigma_i \sigma_j\big) 
+ J_4 \, s_i s_j \sigma_i \sigma_j \Big].
\end{align}

This coupling enables investigation of how different issues interact within complex social discourse. The pairwise interactions ($J_2$) represent direct agreement or disagreement between individuals, while four-spin interactions ($J_4$) capture higher-order correlations emerging from jointly considered issues. For example, one spin variable might represent a preference for a company's phone in consumer behavior. In contrast, the other represents a preference for the same company's laptop when $J_4>0$, a \textit{synergistic effect} emerges where satisfaction with one product enhances preference for the other, reflecting brand loyalty. Conversely, in public policy formation, where resources are limited, one spin variable might represent views on economic development while the other captures environmental conservation attitudes. Here, strong consensus on economic priorities often weakens environmental advocacy---a competitive \textit{antagonism} captured by antiferromagnetic coupling ($J_4<0$). Previous studies have focused primarily on ferromagnetic coupling ($J_2,J_4>0$)~\cite{jang2015ashkin, kim2021link}, where consensus on one issue reinforces agreement on the other. However, real social systems frequently exhibit \textit{competing priorities}, from budget allocations to policy trade-offs, making the antiferromagnetic regime especially relevant.

Our analysis reveals three key findings. First, we show that highly connected nodes (hubs) play a crucial role in determining overall system dynamics by creating strong local spin correlations that promote order. Second, our results indicate that systems with competing interactions maintain their hub-influenced behavior across a significantly wider range of network structures than systems with cooperative interactions. Third, we discover unique asymmetric ordered phases, particularly when specific interaction strengths ($J_2$ and $J_4$) are comparable, demonstrating how network architecture can give rise to novel states rarely observed in lattice systems.

The remainder of this paper is organized as follows. Section~\ref{sec:AF-AT model} introduces the AF-AT model and its mean-field solution on random scale-free networks, extending previous studies of AT models~\cite{jang2015ashkin, kim2021link} to antiferromagnetic regimes. In Section~\ref{sec:phase}, we analyze the rich phase diagram featuring four distinct phases: paramagnetic, Baxter, $\langle \sigma \rangle$, and antiferromagnetic phases, along with their transitions and stability conditions. Section~\ref{sec:lambda-Dependence} examines how network heterogeneity influences phase behavior, particularly focusing on the extended critical regime $5 < \lambda < 9.237$ where hub effects remain significant. Finally, Section~\ref{sec:discussion} explores the implications of our findings for understanding opinion dynamics and consensus formation in social networks.

\section{The AF-AT model}\label{sec:AF-AT model}

\subsection{Model}\label{sec:model}
\begin{figure}
\includegraphics[width=1.0\linewidth]{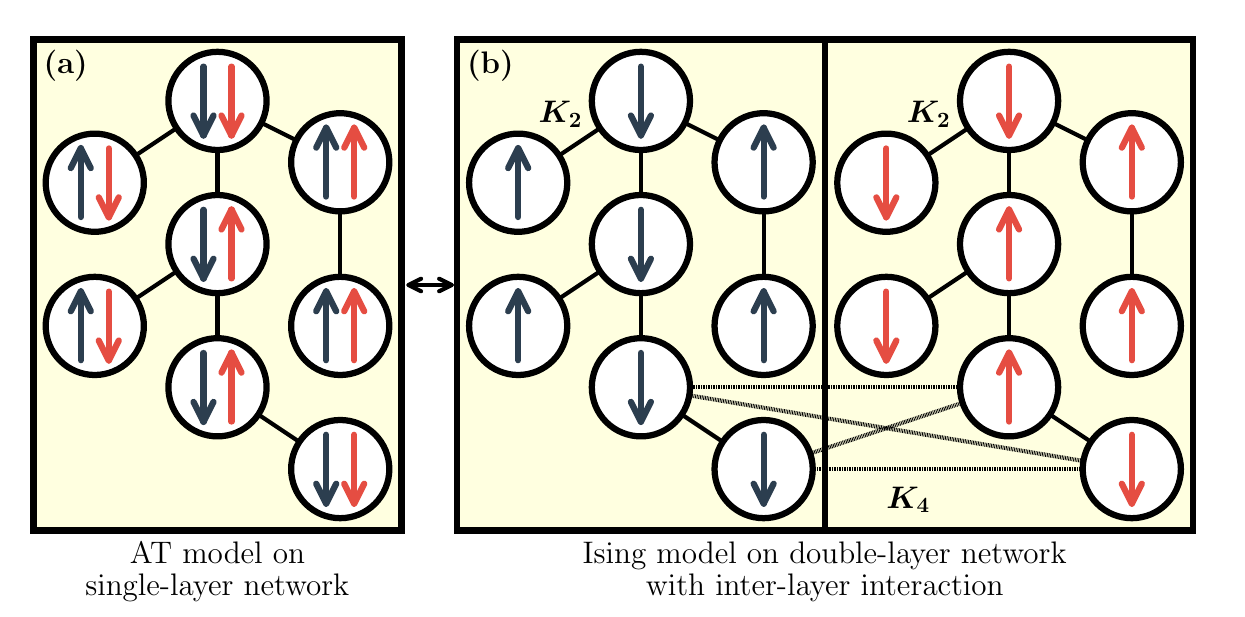}
\caption{
In (a), the Ashkin-Teller (AT) model on a single-layer network shows nodes containing two types of Ising spins (red and blue), with interactions depicted by solid lines. Panel (b) illustrates the equivalent system as a two-layer multiplex network, where each layer contains a single spin type. Solid and dashed lines represent intra- and inter-layer interactions, respectively.
}
\label{fig:fig1}
\end{figure}
The Hamiltonian of the AF-AT model is written as
\begin{equation} \label{eq:hamiltonian}
\mathcal{H} = -\sum_{(i,j)\in\rm{n.n}} \left[J_2(s_is_j + \sigma_i\sigma_j) - J_4 s_i\sigma_i s_j\sigma_j \right],
\end{equation}
where both $J_2$ and $J_4$ are positive.

The AF-AT model has been previously investigated on Euclidean space using mean-field theory~\cite{kohmoto1981hamiltonian}. Here, we study this model on random scale-free (SF) networks, which exhibit heterogeneous degree distributions following a power law $P(k) \simeq k^{-\lambda}$. The degree $k_i$ of node $i$ represents the number of its connected neighbors, while the degree exponent $\lambda$ characterizes network heterogeneity - smaller $\lambda$ indicates more heterogeneous networks due to increased variance in degrees. We consider only $\lambda > 2$, as the mean degree $\langle k \rangle$ diverges for $\lambda < 2$, making such networks physically unrealistic.
Unlike the mean-field solution in Euclidean space, the presence of highly connected nodes (hubs) in SF networks dramatically enriches the phase diagram of the AF-AT model. This hub-mediated spin correlation, which scales proportionally with degree, dramatically alters the system properties compared to homogeneous systems consisting of non-hubs.

\subsection{The free energy}\label{sec:MFT}
Mean-field theory provides a powerful framework for analyzing phase transitions in complex systems. Here, we derive the free energy $\mathcal{F}$ using the Ginzburg-Landau mean-field formalism. This approach gives exact results above the upper critical dimension~\cite{stanley1971phase, kardar2007statistical, goldenfeld2018lectures} and on Erdős-Rényi networks~\cite{lima2012ising}.

\subsubsection{Order parameters and mean-field approximation}
We introduce three local order parameters characterizing the state of each node $i$:
\begin{equation}\label{eq:order_params}
m_{s}^{i} =\langle s_i \rangle, \ m_{\sigma}^{i} =\langle \sigma_i \rangle, \ m_{s\sigma}^i =\langle s_i\sigma_i \rangle,
\end{equation}
where $\langle \cdots \rangle$ denotes the ensemble average. 
{We expand the spin variables around these mean values as:}
\begin{equation}\label{eq:fluctuations}
s_i= m_{s}^{i} + \delta m_{s}^{i}, \ \sigma_i = m_{\sigma}^{i} + \delta m_{\sigma}^{i}, \ s_i\sigma_i=m_{s\sigma}^{i} + \delta m_{s\sigma}^{i}.
\end{equation}

In the mean-field approximation, we neglect fluctuation terms higher than first order, yielding:
\begin{align}\label{eq:mf_hamiltonian}
\mathcal{H} \simeq 
& - J_2 \sum_{(i,j) \in {\rm n.n}} \left(m_s^i m_s^j + m_\sigma^i m_\sigma^j \right) -J_4 \sum_{(i,j) \in {\rm n.n}} m_{s\sigma}^i m_{s\sigma}^j \cr
& - J_2\sum_{(i,j) \in {\rm n.n}} (m_s^j s_i + m_\sigma^j \sigma_i) + J_4 \sum_{(i,j) \in {\rm n.n}} m_{s\sigma}^j s_i \sigma_i,
\end{align}
{where ${\rm n.n}$ is an abbreviation of the nearest neighbor.}

\subsubsection{Network effects and annealed approximation}
To handle heterogeneous networks, particularly random scale-free topologies, we employ the annealed network approximation~\cite{dorogovtsev2002ising, bianconi2002mean, derrida1986random, rohlf2002criticality, leone2002ferromagnetic} as:
\begin{equation}\label{eq:annealed}
\sum_{(i,j) \in {\rm n.n}} \mathcal{A}_{ij} \simeq \frac{1}{2} \sum_{i\neq j} \frac{k_i k_j}{N\langle k\rangle} \mathcal{A}_{ij},
\end{equation}
where $N$ is the network size and $\langle k\rangle = \sum_k kP(k)$ denotes the mean degree. This approximation captures degree heterogeneity while maintaining analytical tractability.

The degree-weighted order parameters are defined as:
\begin{equation}\label{eq:weighted_params}
m_s \equiv \sum_i \frac{k_i m_s^i}{N \langle k\rangle},
\quad m_\sigma \equiv \sum_i \frac{k_i m_\sigma^i}{N\langle k\rangle},
\quad M \equiv \sum_i \frac{k_i m_{s \sigma}^i}{N\langle k\rangle}.
\end{equation}

\subsubsection{Free energy derivation}
The free energy density $f \equiv \mathcal{F} / N$ is obtained through the partition function:
\begin{equation}\label{eq:partition} 
e^{-\beta \mathcal{F}} = \sum_{s, \sigma} e^{-\beta \mathcal{H}},
\end{equation}
where $\beta = 1 / k_B T$. This yields:
\begin{align}\label{eq:fe_density} 
f & \simeq\frac{1}{2}m_s^2 \langle k\rangle / T - \int_1^\infty \ln [\cosh (m_s k / T)] P(k) dk \cr 
& +\frac{1}{2}m_\sigma^2 \langle k\rangle/ T - \int_1^\infty \ln [\cosh (m_\sigma k / T)] P(k) dk \cr 
& -\frac{1}{2}x M^2 \langle k\rangle / T - \int_1^\infty \ln [\cosh (x M k / T)] P(k) dk \cr 
& - \int_1^\infty \ln(\mathcal{B}) P(k) dk, 
\end{align}
{where $x\equiv -J_4/J_2$}, and 
\begin{equation}\label{eq:B_AT} 
\mathcal{B} = 1-\tanh\left(m_s k / T \right) \tanh \left(m_\sigma k / T \right) \tanh \left(x M k / T \right).
\end{equation}
The first six terms represent the energy cost of maintaining the order parameters, while the last term captures their coupling.

\subsubsection{Self-consistency equations}
Minimizing the free energy to the order parameters leads to:
\begin{equation}\label{eq:scr}
\begin{split}
m_s \langle k\rangle        & = \int_1^\infty k C_s  (k) P(k) dk \cr
m_\sigma   \langle k\rangle & = \int_1^\infty k C_\sigma   (k) P(k) dk \cr
M\langle k\rangle & = \int_1^\infty k C_{s\sigma}(k) P(k) dk,
\end{split}
\end{equation}
where
\begin{align}\label{eq:correlations}
C_s(k)        & = \dfrac{\T(m_s k / T)-\T(m_\sigma k / T)\T(x M k / T)}
{1 - \T(m_s k / T)\T(m_\sigma k / T)\T(x M k / T)}, \cr
C_\sigma(k)   & = \dfrac{\T(m_\sigma k / T)-\T(m_s k / T)\T(x M k / T)}
{1 - \T(m_s k / T)\T(m_\sigma k / T)\T(x M k / T)}, \cr
C_{s\sigma}(k)& = \dfrac{-\T(x M k / T)+\T(m_s k / T)\T(m_\sigma k / T)}
{1 - \T(m_s k / T)\T(m_\sigma k / T)\T(x M k / T)},
\end{align}
with $\T(x) \equiv \tanh(x)$. These equations determine the equilibrium values of the order parameters.

\section{Phases, Phase diagram, and Phase transitions}\label{sec:phase}

\begin{figure}   
\includegraphics[width=0.8\linewidth]{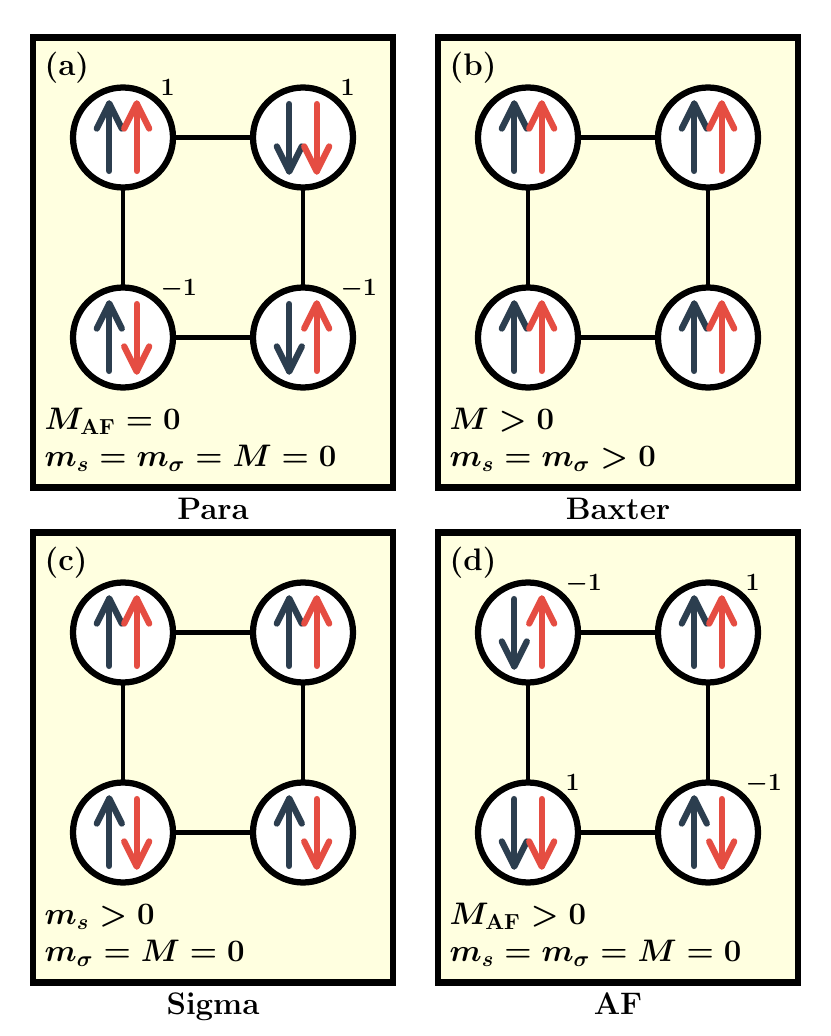}
\caption{Distinct spin alignments characterizing ground states in each phase: paramagnetic (a), Baxter (b), $\langle \sigma \rangle$ (c), and antiferromagnetic (d). Each phase exhibits unique order parameter values, as indicated below each configuration.
}
\label{fig:fig2}	
\end{figure}

Building on the self-consistency equations derived in Sec.~\ref{sec:MFT}, we identify four distinct phases in the AF-AT model. As shown in Fig.~\ref{fig:fig2}, these phases emerge from different spin configurations of the order parameters $m_s$, $m_\sigma$, $M$ and $M_{\AF}$, with particularly interesting behavior in the $\langle \sigma \rangle$ phase. This phase represents a partial ordering where symmetry between two layers is broken—one magnetization vanishes while the other remains finite (here, $m_s = 0$, $m_\sigma \ne 0$) with no cross-correlation ($M=0$). This asymmetry arises from the competition between ferromagnetic intra-layer interactions and antiferromagnetic inter-layer coupling, where strong ordering in one layer actively suppresses order formation in the other. This antagonistic effect between two layers directly results from the antiferromagnetic nature of the four-spin interactions and distinguishes the AF-AT model from conventional coupled spin systems.

\subsection{Phase Characteristics}
(i) Paramagnetic (PM) phase: At high temperatures, thermal fluctuations dominate, and all order parameters vanish ($m_s = m_\sigma = M = 0$), resulting in zero free energy ($f_\PM=0$).

(ii) Baxter phase: This phase exhibits symmetric ordering of both spin types {($m_s=m_\sigma=m>0$)} with positive cross-correlation ($M>0$), analogous to ferromagnetism in coupled spin systems.
\begin{align}\label{eq:f_bx}
& f_{\Baxter}(m, M) = m^2 \kbar / T - 2\int_1^\infty \ln [ \cosh(m k / T) ] P(k) dk \cr
& - \frac{1}{2} x M^2 \kbar / T - \int_1^\infty \ln [\cosh(x M k / T)] P(k) dk \cr
& - \int_1^\infty \ln \left[ 1 - \tanh^2 (m k / T) \tanh (x M k / T) \right] P(k) dk.
\end{align}

(iii) $\langle \sigma \rangle$ phase: This phase exhibits a partial ordering where symmetry between spins is broken—one magnetization vanishes while the other remains finite (here, $m_s=0$, $m_\sigma\neq0$) with no cross-correlation ($M=0$). The asymmetry arises from the competition between ferromagnetic and antiferromagnetic interactions. The free energy is:
\begin{equation}\label{eq:f_sg}
f_{\langle \sigma \rangle} (m_\sigma) = \frac{1}{2} m_\sigma^2 \kbar /T - \int_1^\infty \ln[\cosh(m_\sigma k / T) ] P(k) dk.
\end{equation}

(iv) Antiferromagnetic (AF) phase: While global magnetizations vanish ($m_s = m_\sigma = M = 0$), this phase exhibits local antiferromagnetic order. Neighboring spin products ($s \sigma$) tend to align in opposite directions, creating a staggered pattern quantified by:
\begin{equation}
M_\AF \equiv \sum_i - \text{sgn}\left(M_{s \sigma}^i \right) \sum_{j \in n.n.(i)} \frac{k_j M_{s \sigma}^j } {N \kbar}.
\end{equation}
This order parameter captures the microscopic antiparallel alignment despite zero bulk magnetization. The corresponding free energy is:
\begin{align}\label{eq:f_af}
f_{\AF} (M_{\AF}) 
& = \frac{1}{2} x M_\AF^2 \kbar / T \cr
& - \int_1^\infty \ln[\cosh(x M_\AF k / T)] P(k) dk.
\end{align}

\begin{figure*}
\includegraphics[width=0.84\textwidth]{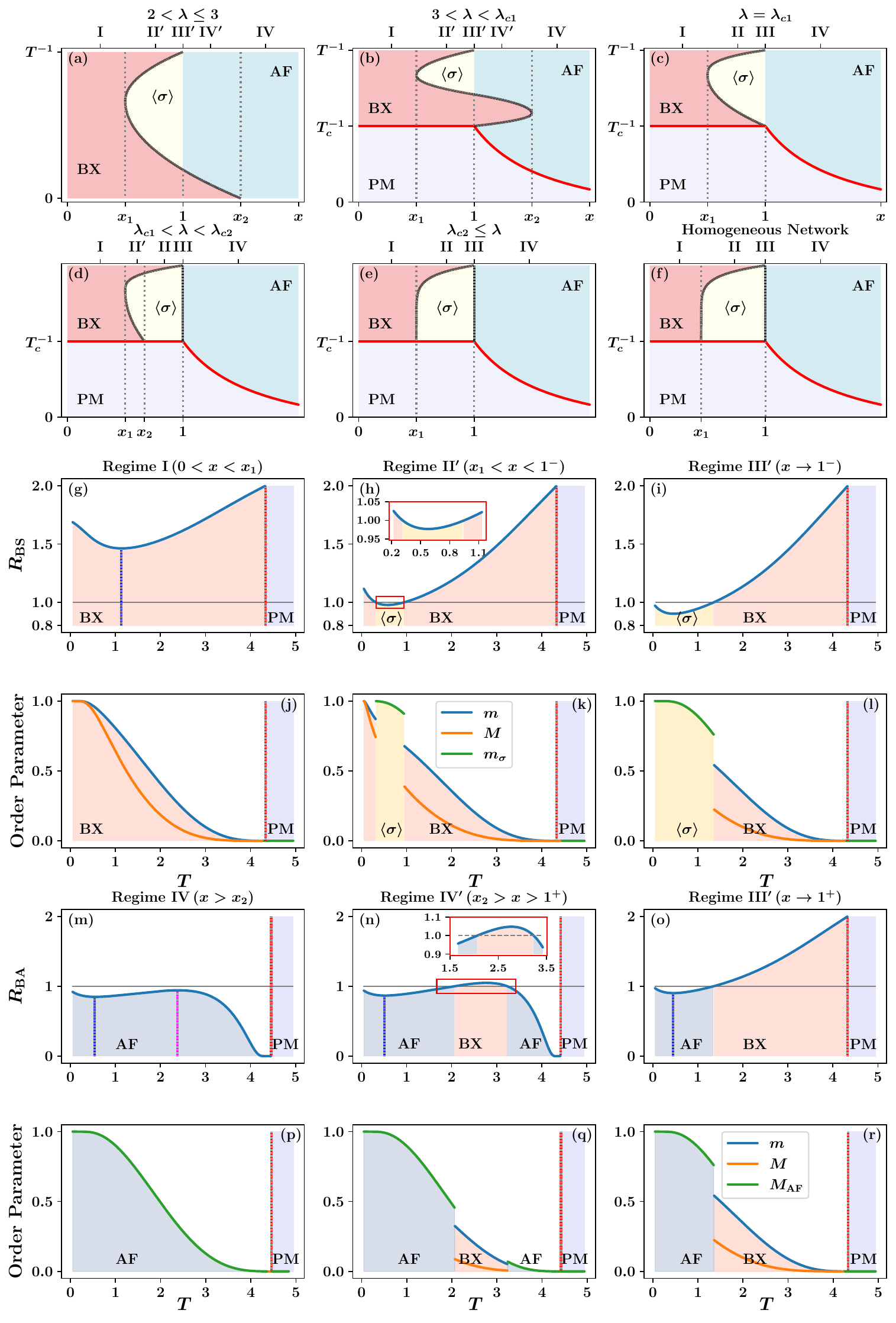}
\caption{Phase behavior of the AF-AT model across different network types. Panels (a-e) display phase diagrams for random scale-free networks with varying degree exponents $\lambda$, while (f) shows the homogeneous network case. The critical temperature for continuous phase transitions is indicated by a red line, while the temperature for discontinuous phase transitions is marked by a black line. We examine the order parameters and free energy ratios for $x<1$ (g-l) and $x\geq1$ (m-r), where the critical temperatures are represented by red vertical lines.}
\label{fig:phase_diagram}\vspace{-0.5cm}
\end{figure*}
\setlength{\textfloatsep}{0pt plus 1.0pt minus 1.0pt}

Fig.\ref{fig:phase_diagram} displays the phase diagrams across distinct heterogeneity regimes characterized by $\lambda$: (a) strongly heterogeneous ($2 < \lambda \le 3$), (b) moderately heterogeneous ($3 < \lambda < \lambda_{c1} \approx 5.732$), (c) critical ($\lambda=\lambda_{c1}$), (d) weakly heterogeneous ($\lambda_{c1} < \lambda < \lambda_{c2} \approx 9.237$), and (e) effectively homogeneous ($\lambda_{c2} \le \lambda$). While conventional spin models exhibit an upper critical degree exponent $\lambda=5$\cite{kohmoto1981hamiltonian}, the AF-AT model extends this to $\lambda_{c2}$, indicating enhanced stability of hub-mediated correlations under antiferromagnetic coupling.

\subsection{Strong Heterogeneity Regime ($2 < \lambda \le 3$)}
In this regime, the strong heterogeneity leads to enhanced spin ordering through hub-mediated spin correlation, suppressing the PM phase entirely. Ordered phases (Baxter, $\langle \sigma \rangle$, AF) persist across all finite temperatures $T > 0$. The free energies expand to the lowest order as:
\begin{equation}
\begin{split}
f_\Baxter (m) & = C^{'}_{23}(\lambda, x, {r\equiv M/m)} [m / T]^{\lambda-1} + O\left(m^2 \right), \cr
f\langle \sigma \rangle (m_\sigma) & = C_{23}(\lambda) [m_\sigma / T]^{\lambda-1} + O\left(m_\sigma^2 \right), \cr
f_\AF (M_\AF) & = C_{23}(\lambda) [x M_\AF / T]^{\lambda-1} + O\left( M_\AF^2 \right),
\end{split}
\end{equation}
The coefficients $C_{23}(\lambda)$ and $C'_{23}(\lambda,x,r)$, {which are defined in Append.~A,} remain negative for all $T$, pushing the transition point to $T=\infty$.
Consequently, the free energies stay negative, suppressing the PM phase entirely. Phase transitions between ordered states occur discontinuously, with the system adopting the phase of lowest free energy. The phase behavior exhibits distinct regimes in coupling strength $x$:

(I) Weak coupling ($x < x_1$): Intra-layer interactions dominate, maintaining nonzero order parameters ($m_s$, $m_\sigma$, $M$) and stabilizing the Baxter phase throughout.

(II$^\prime$) Intermediate coupling ($x_1 < x < 1$): Strong hub effects still suppress PM phase, but increased AF coupling enables $\langle \sigma \rangle$ phase at intermediate temperatures, with discontinuous transitions to Baxter phase [Fig.~\ref{fig:phase_diagram}(a)].

(IV$^\prime$) Strong coupling ($1< x < x_2$): The AF phase dominates at low temperatures but yields the Baxter phase at higher temperatures.

(IV) Very strong coupling ($x > x_2$): AF phase persists across all finite temperatures.

\subsection{Moderately Heterogeneous Regime ($3 < \lambda < \lambda_{c1}$)}
The free energies expand to two lowest orders:
\begin{equation}
\begin{split}
f_{\Baxter} (m)
& =  m^2 \dfrac{\kbar}{T} \left[1-\dfrac{{\kkbar}/{\kbar}}{T}\right] \cr
& + 2 C_{35}(\lambda)[m / T]^{\lambda-1} + O\left( m^{4} \right), \cr
f_{\langle \sigma \rangle} (m_\sigma) & = \frac{1}{2} m_\sigma^2 \dfrac{\kbar}{T} \left[1-\dfrac{{\kkbar}/{\kbar}}{T}\right] \cr
& + C_{35}(\lambda)[m_\sigma / T]^{\lambda-1} + O\left( m_\sigma^{4} \right), \cr
f_\AF (M_\AF)
& = \frac{J}{2} M_\AF^2 \dfrac{\kbar}{T} \left[1 - x\dfrac{{\kkbar}/{\kbar}}{T} \right] \cr
& + C_{35}(\lambda)[x M_\AF / T]^{\lambda-1} + O\left( M_\AF^{4} \right).
\end{split}
\end{equation}
{The coefficient $C_{35}(\lambda)$ is defined in Append.~A.}
These expansions reveal critical transition temperatures: (i) $T_c=\langle k^2 \rangle/\langle k \rangle$ for Baxter and $\langle \sigma \rangle$ phases ($0< x < 1$), and (ii) $T^{'}_{c}=x\langle k^2 \rangle/\langle k \rangle$ for the AF phase ($x > 1$). At these points, both free energy minima and order parameters vanish, indicating continuous transitions to disordered phase (PM) characterized by critical exponents in Table~\ref{tab:exponents}.
\subsubsection{Weak Coupling Regime ($0< x \le 1$)}
For $x < 1$, the system exhibits either Baxter or $\langle \sigma \rangle$ phase below $T_c = \kkbar/\kbar$, determined by the free energy ratio:
\begin{equation}
R_{\BS}(T) \equiv \left| \frac{f_{\Baxter, \min}}{f_{\langle \sigma \rangle, \min}} \right|.
\end{equation}
The behavior of $R_{\BS}(T)$ [Fig.~\ref{fig:phase_diagram}(g)--(i)] defines three distinct regions:
Region I: $R_{\BS}(T) > 1$ throughout $[0, T_c]$, yielding pure Baxter phase with continuous transition to PM at $T_c$.
Regions II$'$ and III$'$: $R_{\BS}(T) < 1$ within $[T_1, T_2]$, enabling $\langle \sigma \rangle$ phase within Baxter domain through discontinuous transitions.
\subsubsection{Strong Coupling Regime ($x \ge 1$)}
For $x > 1$, the system exhibits either Baxter and AF below $T^{'}_{c} = x\kkbar/\kbar$,
determined by the free energy ratio:
\begin{equation}
R_{\BA}(T) \equiv \left\vert \frac{f_{\Baxter, \min}}{f_{\AF, \min}}\right\vert.
\end{equation}
The behavior of $R_{\BA}(T)$ [Fig.~\ref{fig:phase_diagram}(m)--(o)] defines three distinct regions:
Region IV ($x > x_2$): $R_{\BA}(T) < 1$ throughout $[0, T^{'}_{c}]$, yielding pure AF phase with continuous transition to PM at $T_{c1}$.
Regions IV$'$ and III$'$: $R_{\BA}(T) < 1$ within $[T_3, T_4]$, enabling Baxter phase within AF domain through discontinuous transitions.

\subsection{Critical and Near-Critical Regimes}

(a) At $\lambda=\lambda_{c1}$: Hub influence weakens sufficiently to eliminate Baxter phase protrusion above $x=1$, creating tetracritical point at $(x, T^{-1}) = (1, T_c^{-1})$ where all four phases meet.

(b) For $\lambda_{c1} < \lambda < \lambda_{c2}$: Further reduction in hub effects shrinks Baxter domain while stabilizing $\langle \sigma \rangle$-PM phase boundary.

(c) For $\lambda \geq \lambda_{c2}$: Network becomes effectively homogeneous, exhibiting phase behavior identical to random ER networks [Fig.~\ref{fig:phase_diagram}(e,f)].

This hierarchy of phase diagrams demonstrates how network heterogeneity and coupling strength jointly determine the system's ordering behavior, with rich phase transitions emerging from their interplay.


\begin{table}[h]
\begin{normalsize}
\setlength{\tabcolsep}{6pt}
{\renewcommand{\arraystretch}{1.6}
\begin{tabular}{@{\extracolsep{\fill}}ccccccc} 
\hline
\hline
Phase&$\lambda$&$\alpha$&$\beta_\sigma$&$-$
&$\gamma_{\sigma\pm}$&$-$ \cr
\hline
\multirow{2.0}{*}{\makecell[c]{$\langle \sigma \rangle$}}&$3 <\lambda \le 5$&$\frac{\lambda-5}{\lambda-3}$&$\frac{1}{\lambda-3}$&$-$&$1$&$-$\cr
&$\lambda > 5$&$0$&$\frac{1}{2}$&$-$&$1$&$-$\cr
\hline
\hline
Phase&$\lambda$&$\alpha$&$\beta_s$&$\beta_{s\sigma}$&
$\gamma_{s\pm}$&$\gamma_{s\sigma\pm}$ \cr
\hline
\multirow{3.2}{*}{\makecell[c]{\Baxter}}
&$3 <\lambda \le 4$&$\frac{\lambda-5}{\lambda-3}$&$\frac{1}{\lambda-3}$&$\frac{\lambda-2}{\lambda-3}$&$1$&$0$\cr
&$4 <\lambda \le 5$&$\frac{\lambda-5}{\lambda-3}$&$\frac{1}{\lambda-3}$&$\frac{2}{\lambda-3}$&$1$&$0$\cr
&$\lambda > 5$&$0$&$\frac{1}{2}$&$1$&$1$&$0$\cr
\hline
\hline
Phase&$\lambda$&$\alpha$&$\beta_\AF$&$-$
&$\gamma_{\AF\pm}$&$-$ \cr
\hline
\multirow{2.0}{*}{\makecell[c]{\AF}}&$3 <\lambda \le 5$&$\frac{\lambda-5}{\lambda-3}$&$\frac{1}{\lambda-3}$&$-$&$1$&$-$\cr
&$\lambda > 5$&$0$&$\frac{1}{2}$&$-$&$1$&$-$\cr
\hline
\hline
\end{tabular}}
\caption{Critical exponents for continuous phase transitions. Here $\alpha$ is the critical exponent of the specific heat, $\beta_q$ is the critical exponent of the magnetization $M_q$ at zero external magnetic fields, and $\gamma_{q}$ is the critical exponent of the susceptibility for $M_q$-magnetization near the transition temperature, where $q \in {s, \sigma, s\sigma, \AF}$.}
\label{tab:exponents}
\end{normalsize}
\end{table}


\section{Effect of network heterogeneity}
\label{sec:lambda-Dependence}

The phase diagram of the AF-AT model depends crucially on the network heterogeneity, characterized by the degree distribution exponent $\lambda$. We focus here on determining two critical values, $\lambda_{c1}$ and $\lambda_{c2}$, that mark fundamental changes in system behavior.

\begin{figure}
    \resizebox{0.6666666666\columnwidth}{!}{\includegraphics{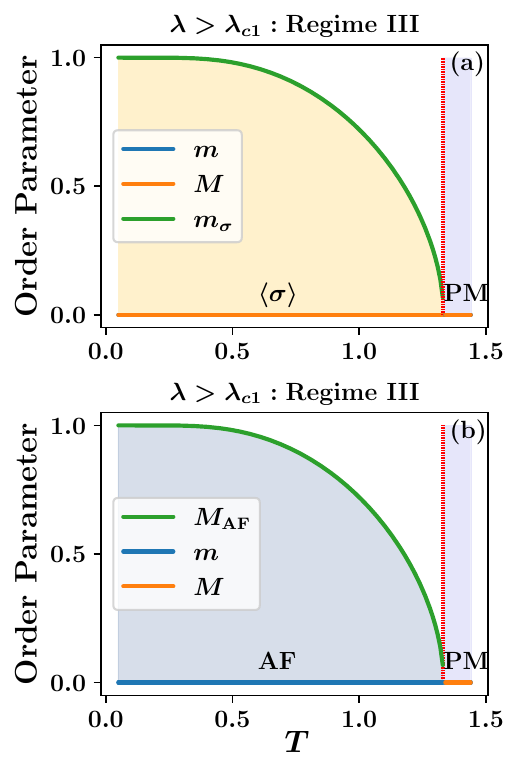}}
    \caption{ (Color online) Temperature dependence of order parameters in Regime III when $\lambda > \lambda_{c1}$. Panel (a) at $x=1^-$ shows the $\langle \sigma \rangle$-PM phase transition with order parameters $m$, $m_{s}$, and $M$. Panel (b) at $x=1^+$ depicts the AF-PM transition characterized by $M_{AF}$, $m$, and $M$. Red dotted lines indicate critical temperatures.
    \label{fig:fig7}	
    } 
\end{figure}

For $\lambda_{c1}$, when $2 < \lambda < \lambda_{c1}$, the Baxter phase extends from Region II$'$ into IV$'$, as shown in Fig.~\ref{fig:phase_diagram}(b), protruding within temperature range [$T_\Baxter$, $T_c$] at $x=1$. As $\lambda$ approaches $\lambda_{c1}^-$, $T_\Baxter$ converges to $T_c^-$. At $\lambda_{c1}$, we find $T_\Baxter=T_c$ at $x=1$, leading to:
\begin{equation}\label{eq:lambda_c1}
1 - \frac{3}{2}\frac{(\lambda_{c1}-3)(\lambda_{c1}-5)}{(\lambda_{c1}-4)^2} = 0.
\end{equation}
The detailed derivation is presented in Appendix.~\ref{seca:lambda-Dependence}.

\begin{figure}[htbp]
    \resizebox{0.6666666666\columnwidth}{!}{\includegraphics{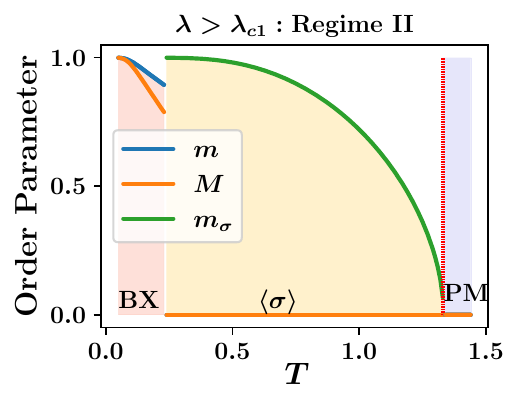}}
    \caption{ (Color online)  Temperature dependence of order parameters in Regime II at $(\lambda, x)=(6, 0.8)$. As temperature increases, the system undergoes successive transitions through Baxter, $\langle \sigma \rangle$, and PM phases. This regime's presence becomes more pronounced in the $\lambda > \lambda_{c1}$ phase diagram. A red dotted line marks the critical temperature.
    \label{fig:fig8}	
    } 
\end{figure}

When $\lambda > \lambda_{c1}$, the Baxter phase is confined to the region $x < 1$ (weak coupling regime) and separated from the AF phase shown in Fig.~\ref{fig:phase_diagram}(d). Regime III$'$ changes to Regime III (Fig.~\ref{fig:fig7}), and a new Region II (Fig.~\ref{fig:fig8}) emerges between Regime II$^{'}$ and III.

For $\lambda_{c2}$, when $\lambda_{c1} < \lambda < \lambda_{c2}$, the boundary between the Baxter phase and the $\langle \sigma \rangle$ phase has a finite slope as shown in Fig.~\ref{fig:phase_diagram}(d) Regime II$^{'}$. In this case, the gradient of $R_{\BS}$ maintains a positive value near the critical temperature in II$^{'}$ (see Fig.~\ref{fig:phase_diagram}(h)). As $\lambda$ approaches $\lambda_{c2}$, the gradient of $R_{\BS}$ becomes zero at $(x, T) = (x_1, T_c)$ as Regime II$^{'}$ shrinks and vanishes. At $\lambda_{c2}$, we find $dR_{\rm BS}(T)/dT=0$ at $(x, T) = (x_1, T_c)$, leading to: \begin{align}\label{eq:BX_COP_condition}
2 \lambda_{c2}^4 -42 \lambda_{c2}^3 + 315 \lambda_{c2}^2 - 1055 \lambda_{c2} + 1410 = 0.
\end{align}
The detailed derivation of the above equation is presented in Appendix.~\ref{seca:lambda-Dependence}.

When $\lambda > \lambda_{c2}$, $R_{\BS}$ decreases monotonically with increasing temperature ($dR_{\rm BS}(T)/dT<0$), reaching unity as its minimum value at $(x, T) = (x_1, T_c)$. As a result, the $\langle \sigma \rangle$ phase first emerges at $(x, T) = (x_1, T_c)$, and the boundary between the Baxter phase and the $\langle \sigma \rangle$ phase is parallel to the temperature axis as shown in Fig.~\ref{fig:phase_diagram}(e). 

\begin{figure}
    \resizebox{0.6666666666\columnwidth}{!}{\includegraphics{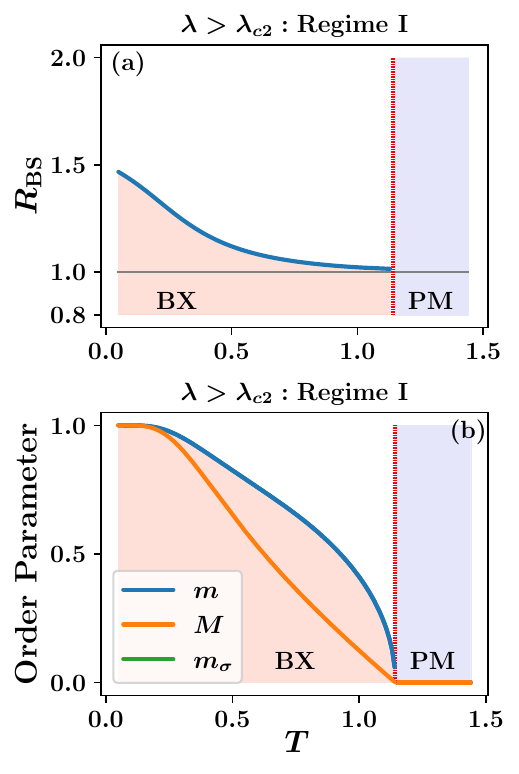}}
    \caption{ (Color online) Behavior of the system in Regime I at $(\lambda, x)=(10, 0.5)$ when $\lambda > \lambda_{c2}$. Panel (a) shows the monotonic decrease of the free energy ratio $R_{\BS}(T)$, while panel (b) displays the temperature dependence of order parameters through the Baxter-PM transition. Critical temperature marked by red dotted lines.
    \label{fig:fig9}	
    } 
\end{figure}

Therefore, $\lambda$ exceeds $\lambda_{c2}$, network heterogeneity becomes irrelevant, and the phase diagram becomes equivalent to that of homogeneous networks. This extension of the critical degree exponent ($\lambda_c = 5 < \lambda_{c2} \simeq 9.237$) highlights that, under antagonistic interlayer effects, network topology exerts a stronger influence on phase behavior. Consequently, hub-mediated spin correlations remain significant across a broader heterogeneity spectrum than in conventional ferromagnetic systems.

\section{Discussion and Conclusion}\label{sec:discussion}

In this study, we have developed a comprehensive theoretical framework to understand phase transitions in the antiferromagnetic Ashkin--Teller (AF-AT) model on random scale-free networks. Using a mean-field analysis, we uncovered a remarkably rich phase diagram that includes Baxter, $\langle \sigma \rangle$, antiferromagnetic, and paramagnetic phases. By rigorous examination of the system's phase diagram, we found that the upper critical degree exponent reaches $\lambda \approx 9.237$, significantly exceeding the conventional threshold $\lambda = 5$ observed in ferromagnetic systems. This striking discrepancy demonstrates how antiferromagnetic interactions can fundamentally reshape the impact of network heterogeneity.

Hub nodes play a central role in shaping the system's behavior. For $2 < \lambda < 3$, hub-dominated local correlations eliminate the paramagnetic phase entirely by driving their neighborhoods toward coordinated states. Their dense connectivity with neighboring nodes stabilizes the Baxter phase and extends its persistence in the parameter space, even when antiferromagnetic interactions typically favor $\langle \sigma \rangle$ or antiferromagnetic order. Through these highly connected hubs, strong local correlations significantly influence the system's macroscopic properties, underscoring how network heterogeneity fundamentally alters overall behavior.

The impact of such network heterogeneity crucially depends on the nature of interactions. Whereas homogeneous behavior emerges at moderate degree exponents ($\lambda_c = 5$)~\cite{dorogovtsev2002ising, igloi2002first, leone2002ferromagnetic, bianconi2002mean, herrero2004ising, lee2009critical} in ferromagnetic models, antiferromagnetic interactions maintain significant local correlations up to $\lambda_{c} \simeq 9.237$. This substantial gap highlights the necessity of jointly considering network topology and specific interaction types when analyzing critical phenomena.

Another notable observation is the emergence of the $\langle \sigma \rangle$ phase across a broad region of the phase diagram, particularly near $x = 1$ where $J_2 \approx J_4$. Whereas this phase is absent in two-dimensional lattices~\cite{ditzian1980phase} and only narrowly appears in three-dimensional systems~\cite{kohmoto1981hamiltonian}, it manifests prominently in random scale-free networks. In this phase, a finite $\langle \sigma \rangle$ coexists with a vanishing $\langle s \rangle$, demonstrating how a dominant opinion or state in one domain can suppress consensus in another. Such a phenomenon might correspond, for instance, to a social network scenario where an overwhelmingly popular stance on one issue diminishes collective engagement on other issues.

Our results, therefore, have important implications for social systems featuring competing forces that shape complex patterns of consensus and discord. Real-world scenarios may include debates between development and conservation agendas or public health measures versus conspiracy theories. By refining our understanding of how antiferromagnetic interactions operate within heterogeneous networks, this framework provides valuable insights into polarization and consensus formation dynamics.

Future directions of study are twofold. First, while we have examined antiferromagnetic inter-layer interactions, we need to consider more generalized forms of spin interactions, including ferromagnetic, antiferromagnetic, and mixed couplings both within and between layers. The competition between different types of spin interactions could reveal novel phase diagrams that may help explain the complex and diverse nature of opinion dynamics in human society. Second, our approach should be extended to more diverse network architectures, where we can systematically investigate how modified spin couplings change the influence of network topology or heterogeneity on macroscopic properties. This expansion is particularly motivated by our finding of an unexpectedly large critical exponent ($\lambda \approx 9.237$), which may open new perspectives on the interplay between spin interactions and network heterogeneity in shaping polarization and consensus formation. Future studies involving modifications of the non-equilibrium voter model \cite{castellano2009statistical} toward the antiferromagnetic AT spin model are promising.

During the preparation of this work the author(s) used [Claude 3.5 Sonnet] and [ChatGPT o1] in order to [language clarity and readability]. After using this tool/service, the author(s) reviewed and edited the content as needed and take(s) full responsibility for the content of the publication.

\begin{acknowledgments}
B.K. was supported by the National Research Foundation of Korea by Grant No. RS-2023-00279802 and the KENTECH Research Grant No. KRG-2021-01-007.
\end{acknowledgments}


\bibliographystyle{apsrev4-2}

%


\clearpage

\newpage

\appendix

\onecolumngrid

\section{Definitions for coefficients}
\label{seca:coeff}
First, the coefficients $\{C_i\}$ are defined as follows:
\begin{align}
    C_{23} (\lambda)
    & = N_\lambda \int_0^\infty \left[ -\ln (\cosh y) \right] y^{-\lambda},\cr
    C_{35} (\lambda)
    & = N_\lambda \int_0^\infty \left[ -\ln (\cosh y) + \frac{1}{2} y^2 \right] y^{-\lambda},\cr
    C_{57} (\lambda)
    & = N_\lambda \int_0^\infty \left[ -\ln (\cosh y) + \frac{1}{2} y^2 - \frac{1}{12} y^4 \right] y^{-\lambda}, \cr
    C_{2}^{(0)} (\lambda, x, r)
    & = N_\lambda \int_0^\infty \left[-\log\left(1-\tanh(xry)[\tanh(y)]^2\right)\right] y^{-\lambda} dy, \cr
    C_{2}^{(1)} (\lambda, x, r)
    & \equiv \frac{\partial}{\partial r} C_{2}^{(0)} (\lambda, r) \cr
    & = N_\lambda \int_0^\infty -\left\{\dfrac{ - [\tanh(y)]^2 + [\tanh(xry)]^2[\tanh(y)]^2}{1-\tanh(xry)[\tanh(y)]^2} \right\} y^{-\lambda} dy, \cr
    C_{23}^{\prime} (\lambda, x, r)
    & = \left[2 + (xr)^{\lambda-1}\right] C_{23}(\lambda) + C_{2}^{(0)}(\lambda, x, r), \cr
    C_{3} (\lambda)
    & = N_\lambda \int_0^\infty \left([\tanh(y)]^2 \right) y^{1-\lambda} dy, 
\end{align}
where $N_\lambda = \lambda - 1$, $r = m/M$.
And next, the coefficients $D_i$ and $E_i$ are defined as follows:
\begin{align}
    D 
    & = \frac{1 + 4 x \kbar / T}{6(1 + x \kbar / T)} \cr
    D_{34} (\lambda) 
    & = \frac{x K_2 [(\lambda-1) C_{3}]^2}{2 \kbar \left[1 + x (\kkbar/\kbar) / T \right]}, \cr
    D_{45} (\lambda)
    & = \frac{1}{6} \left[ \frac{\lambda-1}{\lambda-5} + \frac{3 x K_2 \langle k^3 \rangle\, \langle k^3 \rangle}{\kbar(1 + x K_2 \kkbar / \kbar)} \right], \cr
    D_{5} (\lambda)
    & = \frac{1}{6} \left[ 1 + \frac{3 x K_2\langle k^3 \rangle\, \langle k^3 \rangle}{\langle k^4 \rangle\,\kbar(1 + x K_2 \kkbar / \kbar)} \right] \langle k^4 \rangle, \cr
    E 
    & = - \frac{2 + 32 x \kbar / T}{45(1 + x \kbar / T)} \cr
    E_{57} (\lambda)
    & = \frac{2}{45} \left[ 1 + \frac{15 x K_2\langle k^5 \rangle\, \langle k^3 \rangle}{\langle k^6 \rangle\,\kbar(1 + x K_2 \kkbar / \kbar)} \right] \langle k^6 \rangle,
\end{align}

\clearpage
\newpage

\section{Series Expansion of the Free Energy $f$}
\label{seca:Free_Energy_f}

\subsection{near the absolute zero temperature $T_0$}

Near $T_0$, the free energy $f$ has a local minimum very close to $1$, so all the order parameters that satisfy the self-consistency relations should be approximated as $1 - \epsilon$ ($\epsilon \ll 1$). 
Since higher-order terms of $m/T$ cannot be neglected, expanding $f$ in terms of $m/T$ is invalid.
Near $T_0$, since $\exp{(-\kbar / T)}$ becomes substantially small, the free energy $f$ is expanded in terms of $\exp{(-\kbar / T)}$.

\subsubsection{random scale-free networks with $\lambda > 2$}
Near $T_0$, the free energies are expanded in terms of $\exp{(-\kbar / T)}$:
\begin{align}
    f_{\Baxter} 
    & = - (1 - \frac{x}{2})\kbar / T - \mathcal{O}\left(\exp(-2 (1 - x) \bar{k} / T)\right)\,, \cr
    f_{\langle \sigma \rangle}
    &= - \frac{1}{2} \kbar / T -\log2 - \mathcal{O}\left(\exp(-2 \bar{k} / T)\right)\,, \cr
    f_{\AF}
    &= - \frac{1}{2} x \kbar / T -\log2 - \mathcal{O}\left(\exp(2 x \bar{k}/ T)\right)\,.
    \label{eq:freeEnergyDensity_AT_0_SF}
\end{align}

\subsubsection{homogeneous networks}
Near $T_0$, the free energies expand in terms of $\exp{(-\kbar / T)}$:
\begin{align}
    f_{\Baxter}
    & = - (1 - \frac{x}{2})\kbar / T - \mathcal{O}\left(\exp(-2 (1 - x) \kbar / T)\right)\,, \cr
    f_{\langle \sigma \rangle}
    &= - \frac{1}{2} \kbar / T - \log2 - \mathcal{O}\left(\exp(-2 \kbar / T)\right)\,, \cr
    f_{\AF} 
    &= -\frac{1}{2} x \kbar / T -\log2 - \mathcal{O}\left(\exp(2 x \kbar / T)\right)\,.
    \label{eq:freeEnergyDensity_AT_0_LMF}
\end{align}
The term $\log 2$ in $f_{\langle \sigma \rangle}$ and $f_{\AF}$ emerges from the $2^N$ states of the disordered order parameters $m_s$ and $m_\sigma$.

\subsection{near the critical temperature $T_c$}

Near the critical temperature $T_c$, the local minimum of $f$ approaches zero, and all order parameters satisfying self-consistency relations become much smaller than one.
Neglecting higher-order terms, we expand $f$ to the three lowest-order terms to the order parameter $M$.

\subsubsection{random scale-free networks with $2<\lambda<3$}
For random scale-free networks with $2<\lambda<3$, $T_c$ diverges and $f$'s local minimum approaches zero as $T \to T_\infty$.
Near $T_\infty$, $f$s are expanded as:
\begin{align}
    f_{\Baxter} (M)
    & = {C_{23}^{\prime}(\lambda,x,r)} [m / T]^{\lambda-1} + m^{2} \kbar / T - \frac{1}{2} x r^2 m^{2} \kbar/ T + \mathcal{O}(m^2)\,, \cr
    f_{\langle \sigma \rangle} (m_{\sigma}) 
    & = {C_{23}(\lambda)} [m_\sigma / T]^{\lambda-1} + \frac{1}{2} m_\sigma^{2} \kbar / T + \mathcal{O}(m_{\sigma}^2)\,, \cr
    f_{\AF} (M_{\AF}) 
    & = {C_{23}(\lambda)} [x M_\AF / T]^{\lambda-1} + \frac{1}{2} x M_\AF^{2} \kbar / T + \mathcal{O}(M_\AF^2)\,.
    \label{eq:freeEnergyDensity_AT_C_SF_2_3}
\end{align}
Here $r$ is defined as, $r \equiv m / M$, derived from the following self-consistency relation,
\begin{align}
     \dfrac{\partial}{\partial M} f_\Baxter(m, M) = 0 \to - x \kbar + (\lambda - 1) C_{23}(\lambda) [(x r)]^{\lambda - 2} + C_{2}^{(1)}(\lambda, x, r) = 0\,,
\end{align}
where the coefficients $C_{23}(\lambda)$, $C_{23}^{\prime}(\lambda,x,r)$, $C_{2}^{(0)}(\lambda,x,r)$, and $C_{2}^{(1)}(\lambda,x,r)$ are given in Append.~\ref{seca:coeff}.

\subsubsection{random scale-free networks with $3<\lambda<4$}
Near $T_c$, the free energies expand as:
\begin{align}
    f_{\Baxter} (m)
    & = m^2 \kbar / T \left[1-\dfrac{({\kkbar} / {\kbar})}{T}\right] + 2 {C_{35}(\lambda)} [m / T]^{\lambda-1}  + D_{34}(\lambda) [m / T]^{2(\lambda-2)} + \mathcal{O}(m^4)\,, \cr
    f_{\langle \sigma \rangle} (m_{\sigma}) 
    & = \frac{1}{2} m_\sigma^2 \kbar /T \left[1-\dfrac{({\kkbar} / {\kbar})}{T}\right] + {C_{35}(\lambda)} [m_\sigma / T]^{\lambda-1} + \mathcal{O}(m_\sigma^4)\,, \cr
    f_{\AF} (M_{\AF}) 
    & = \frac{1}{2} x M_\AF^2 \kbar / T \left[1 - \dfrac{(x {\kkbar} / {\kbar})}{T}\right] + {C_{35}(\lambda)} [x M_\AF / T]^{\lambda-1} + \mathcal{O}(M_\AF^4)\,,
    \label{eq:freeEnergyDensity_AT_C_SF_3_4}
\end{align}
where the coefficients $C_{35}$ and $D_{34}$ are given in Append.~\ref{seca:coeff}. 
To expand $f_\Baxter$ in terms of $m$, here we exploit a relation between $m$ and $M$ given as,
\begin{align}
    M = \frac{C_{3}(\lambda)}{\kbar [1 + x (\kkbar/ \kbar) / T] } [m / T]^{\lambda-2} + \mathcal{O}(m^{\lambda-2})\,.
    \label{eq:relation_M_m_AT_SF_3_4}
\end{align}
This relation is derived from the following equation, which can be obtained by expanding the Eq.~\eqref{eq:scr},
\begin{align}
    \dfrac{\partial}{\partial M} f_\Baxter(m, M) = 0 \to M \kbar \left[1 + x ({\kkbar} / {\kbar} ) / T \right] = {C_{23}(\lambda)} [K_4 M]^{\lambda-2} + {C_{3}(\lambda)} [m / T]^{\lambda-2} + \mathcal{O}(M^2)\,.
\label{eq:equationOfState_M_AT_expansion_3_4}
\end{align}
In Eq.~\eqref{eq:equationOfState_M_AT_expansion_3_4}, $(1-xT/T_c)$ has order $1$, making the left side $\mathcal{O}(M)$.
Near $T_c$, the magnitude of $M$ is much less than $1$; therefore, the first term $M^{\lambda-2}$ on the right side cannot be in the same order as the left. 
Instead, the second term $m^{\lambda-2}$ on the right side should be in the same order as the left side as the Eq.~\eqref{eq:relation_M_m_AT_SF_3_4}.

\subsubsection{random scale-free networks with $\lambda > 4$}
A relation between $M$ and $M$ in the Baxter phase is given as,
\begin{align}
    M = \frac{\langle k^3 \rangle}{\kbar\left[1 + x (\kkbar/\kbar) / T \right]} [m / T]^{2} + \mathcal{O}(m^2)\,,
    \label{eq:relation_M_m_AT_SF_4}
\end{align}
from the following self-consistency relation,
\begin{align}
    \dfrac{\partial}{\partial M} f_\Baxter (m, M) = 0 \to M \kbar \left[1 + x ({\kkbar} / {\kbar} ) / T \right] = {C_{23}(\lambda)} [K_4 M]^{\lambda-2} + \langle k^3 \rangle [m / T]^{2} + \mathcal{O}(M^4)\,.
\label{eq:equationOfState_M_AT_expansion_4}
\end{align}
Using this relation, we expand $f_{\Baxter} (M)$ as follows.

First, for random scale-free networks $4<\lambda<5$, $f$s are expanded as:
\begin{align}
    f_{\Baxter} (m) 
    & = m^2 \kbar / T \left[1-\dfrac{({\kkbar} / {\kbar})}{T}\right] + 2 {C_{35}(\lambda)} [m / T]^{\lambda-1} + {D_{45}(\lambda)} [m / T]^{4} + \mathcal{O}(m^4)\,, \cr
    f_{\langle \sigma \rangle} (m_{\sigma}) 
    & = \frac{1}{2} m_\sigma^2 \kbar /T \left[1-\dfrac{({\kkbar} / {\kbar})}{T}\right] + {C_{35}(\lambda)} [m_\sigma / T]^{\lambda-1} + \frac{1}{12}[m_\sigma / T]^{4} \frac{\lambda-1}{\lambda-5} + \mathcal{O}(m_\sigma^4)\,, \cr
    f_{\AF} (M_{\AF}) 
    & = \frac{1}{2} x M_\AF^2 \kbar / T \left[1-\dfrac{(x {\kkbar} / {\kbar})}{T}\right] + {C_{35}(\lambda)} [x M_\AF / T]^{\lambda-1} + \frac{1}{12}[x M_{\AF} / T]^{4} \frac{\lambda-1}{\lambda-5} + \mathcal{O}(M_\AF^4)\,,
    \label{eq:freeEnergyDensity_AT_SF_4_5}
\end{align}
where coefficients $C_{35}$ and $D_{45}$ are given in Append.~\ref{seca:coeff}.

And next, for random scale-free networks $5<\lambda<7$, $f$s are expanded as:
\begin{align}
    f_{\Baxter} (m) 
    & = m^2 \kbar / T \left[1-\dfrac{({\kkbar} / {\kbar})}{T}\right] + D_{5} [m / T]^{4} + 2 {C_{57}(\lambda)}[m / T]^{\lambda-1} + \mathcal{O}(m^6)\,, \cr
    f_{\langle \sigma \rangle} (m_{\sigma}) 
    & = \frac{1}{2} m_\sigma^2 \kbar /T \left[1-\dfrac{({\kkbar} / {\kbar})}{T}\right] + \frac{1}{12} [M / T]^4 \langle k^4 \rangle + {C_{57}(\lambda)}[m_\sigma / T]^{\lambda-1} + \mathcal{O}(m_\sigma^6)\,, \cr
    f_{\AF} (M_{\AF})
    & = \frac{1}{2} x M_\AF^2 \kbar / T \left[1-\dfrac{(x {\kkbar} / {\kbar})}{T}\right] + \frac{1}{12} [x M_{\AF} / T]^4 \langle k^4 \rangle + {C_{57}(\lambda)}[x M_\AF / T]^{\lambda-1} + \mathcal{O}(M_\AF^6)\,,
    \label{eq:freeEnergyDensity_AT_SF_5_7}
\end{align}
where coefficients $C_{35}$ and $D_{5}$ are given in Append.~\ref{seca:coeff}.

Finally, for random scale-free networks $\lambda>7$, $f$s are expanded as:
\begin{align}
    f_{\Baxter} (m)
    & = m^2 \kbar / T \left[1-\dfrac{({\kkbar} / {\kbar})}{T}\right] + {D_{5}(\lambda)} [m / T]^{4} - {E_{57}(\lambda)} [m / T]^{6} \langle k^6 \rangle + \mathcal{O}(m^6),\cr
    f_{\langle \sigma \rangle} (m_{\sigma}) 
    & = \frac{1}{2} m_\sigma^2 \kbar /T \left[1-\dfrac{({\kkbar} / {\kbar})}{T}\right] + \frac{1}{12} [m_\sigma / T]^4 \langle k^4 \rangle - \frac{1}{45}[m_\sigma / T]^{6} \langle k^6 \rangle + + \mathcal{O}(m_\sigma^6)\,, \cr
    f_{\AF} (M_{\AF}) 
    & = \frac{1}{2} x M_\AF^2 \kbar / T \left[1-\dfrac{(x {\kkbar} / {\kbar})}{T}\right] + \frac{1}{12} [x M_{\AF} / T]^4 \langle k^4 \rangle - \frac{1}{45}[x M_\AF / T]^{6} \langle k^6 \rangle + + \mathcal{O}(M_\AF^6)\,,
    \label{eq:freeEnergyDensity_AT_SF_7}
\end{align}
where coefficients $C_{57}$, $D_{5}$, and $E_{7}$ are given in Append.~\ref{seca:coeff}.

\subsubsection{Homogeneous networks}
Near $T_c$, $f$s is expanded as:
\begin{align}
    f_\Baxter 
    & = m^2 \kbar / T \left(1-\kbar / T\right) + D(m \kbar / T)^{4} + E(m \kbar / T)^{6} + \mathcal{O}(m^6)\,, \cr
    f_{\langle \sigma \rangle} 
    & = \frac{1}{2} m_\sigma^2 \kbar /T \left(1-\kbar / T\right) + \frac{1}{12}(m_\sigma \kbar / T)^{4} - \frac{1}{45}(m_\sigma \kbar / T)^{6} + \mathcal{O}(m_\sigma^6)\,, \cr
    f_{\AF} (M_{\AF}) 
    & = \frac{1}{2} x M_\AF / T^2 \kbar \left(1 - x \kbar / T\right) + \frac{1}{12}(x M_\AF \kbar / T)^{4} + \mathcal{O}(M_\AF^6)\,,
    \label{eq:freeEnergyDensity_AT_C_LMF}
\end{align}
where the coefficient $D$ and $E$ are given in Append.~\ref{seca:coeff}.
To expand $f$ of the Baxter phase as the function of $m$, we use the relation between $m$ and $M$ as
\begin{align}
	M = \frac{1}{1 + x \kbar / T} (m \kbar / T)^{2} + \mathcal{O}(m^2)\,.
	\label{eq:relation_M_m_AT_LMF}
\end{align}
While this relation resembles the random scale-free networks case with $3<\lambda<5$, there is a key difference.
In the random scale-free networks with $3<\lambda<5$, $M$ scales with $\mathcal{O}(m^{\lambda-2})$, while in the homogeneous networks, it scales with $\mathcal{O}(m^{2})$. 
This distinction leads to behaviors near the critical temperature $T_{c}$.

\clearpage
\newpage

\section{$R_{\BS}(T)$ Behavior}
\label{seca:R_function_weak}

The behavior of $R_{\BS}(T)$ reveals distinct characteristics across different network types and temperature regimes. This section analyzes these behaviors near absolute zero ($T_0$) and critical temperature ($T_c$).

\subsection{For random scale-free networks with $3<\lambda<4$}

In random scale-free networks with $3<\lambda<4$, the ratio $R_{\BS}(T)$ exhibits non-monotonic temperature dependence, leading to unconventional phase transitions.

\subsubsection{Behavior near $T_0$}
Near $T_0$, $R_{\BS}(T)$ can be expressed as:
\begin{align}
   R_{\BS}(T) 
   & = \frac{f_{\Baxter, \min}}{f_{\langle \sigma \rangle, \min}} = \frac{|f_{\Baxter, \min}|}{|f_{\langle \sigma \rangle, \min}|} \simeq \frac{2 - x}{1 + 2 T \log 2 / \kbar} + \text{H.O.}, 
   \label{eq:freeEnergyDensity_AT_0_ratio}
\end{align}

\subsubsection{Behavior near $T_c$}
As temperature approaches $T_c$, $R_{\BS}(T)$ takes the form:
\begin{align}
   R_{\BS}(T) 
   & = \frac{f_{\Baxter, \min}}{f_{\langle \sigma \rangle, \min}} = \frac{|f_{\Baxter, \min}|}{|f_{\langle \sigma \rangle, \min}|} \simeq 2 - \frac{2 D}{(\lambda-1)(\lambda-3)C_{1}^2 \kkbar} |t| + \text{H.O.},
   \label{eq:freeEnergyDensity_AT_C_ratio_SF}
\end{align}
where $t = (T - T_c)$. 

The derivation of Eq.~\eqref{eq:freeEnergyDensity_AT_C_ratio_SF} requires determining the local minimum of $f$:
\begin{align}
   m^{*} / T \simeq(\frac{|t|}{(\lambda - 1) C_{35} \kbar / T})^{1/(\lambda-3)} + \text{H.O.}.
   \label{eq:order_parameter_AT_C_SF}
\end{align}

Eq.~\eqref{eq:order_parameter_AT_C_SF} characterizes order parameters in both the Baxter and $\langle \sigma \rangle$ phases.

\subsubsection{Temperature dependence implications}
The value of $R_{\BS}(T)$ exhibits non-monotonic behavior. This non-monotonic behavior induces a phase transition from the $\sigma$ phase to the Baxter phase, as demonstrated in [Fig.~\ref{fig:phase_diagram}(j)--(l)].

\subsection{Homogeneous networks}

For homogeneous networks, $R_{\BS}(T)$ demonstrates a simpler, monotonic behavior that contrasts with the random scale-free network case.

\subsubsection{Behavior near $T_0$}
Near $T_0$, $R_{\BS}(T)$ can be expressed as:
\begin{align}
   R_{\BS}(T) 
   & = \frac{f_{\Baxter, \min}}{f_{\langle \sigma \rangle, \min}} = \frac{|f_{\Baxter, \min}|}{|f_{\langle \sigma \rangle, \min}|} \simeq \frac{2 - x}{1 + 2 T \log 2 / \kbar} + \text{H.O.}, 
   \label{eq:freeEnergyDensity_AT_0_ratio_LMF}
\end{align}

\subsubsection{Behavior near $T_c$}
As temperature approaches $T_c$, $R_{\BS}(T)$ takes the form:
\begin{align}
   R_{\BS}(T) = \frac{f_{\Baxter, \min}}{f_{\langle \sigma \rangle, \min}} = \frac{|f_{\Baxter, \min}|}{|f_{\langle \sigma \rangle, \min}|} \simeq 2\frac{1 + x}{1 + 4 x} -  F |t| + \text{H.O.}, \quad F = x\frac{7 - 8x}{30 (-1 + 4 x)^3}
   \label{eq:freeEnergyDensity_AT_C_ratio_LMF}
\end{align}
The coefficient $F$ maintains positive values within $0 \le x < 7/8$, including Regime I ($[0, 1/2]$ as shown in [Fig.~\ref{fig:phase_diagram}(f)]).

To derive Eq.~\eqref{eq:freeEnergyDensity_AT_C_ratio_LMF}, we determine the local minimum of $f$:
\begin{align}
   \kbar  m^{*} / T \simeq(- \frac{3 |t|}{\kbar / T}\frac{1 + x}{1 + 4 x})^{1/2} \quad \text{ for Baxter }, \quad \kbar m^{*} / T \simeq(- \frac{3 |t|}{\kbar / T})^{1/2} \quad \text{for $\langle \sigma \rangle$}.
   \label{eq:order_parameter_AT_C_LMF}
\end{align}

\subsubsection{Temperature dependence implications}
\begin{figure}[ht]
    \resizebox{1.0\columnwidth}{!}{\includegraphics{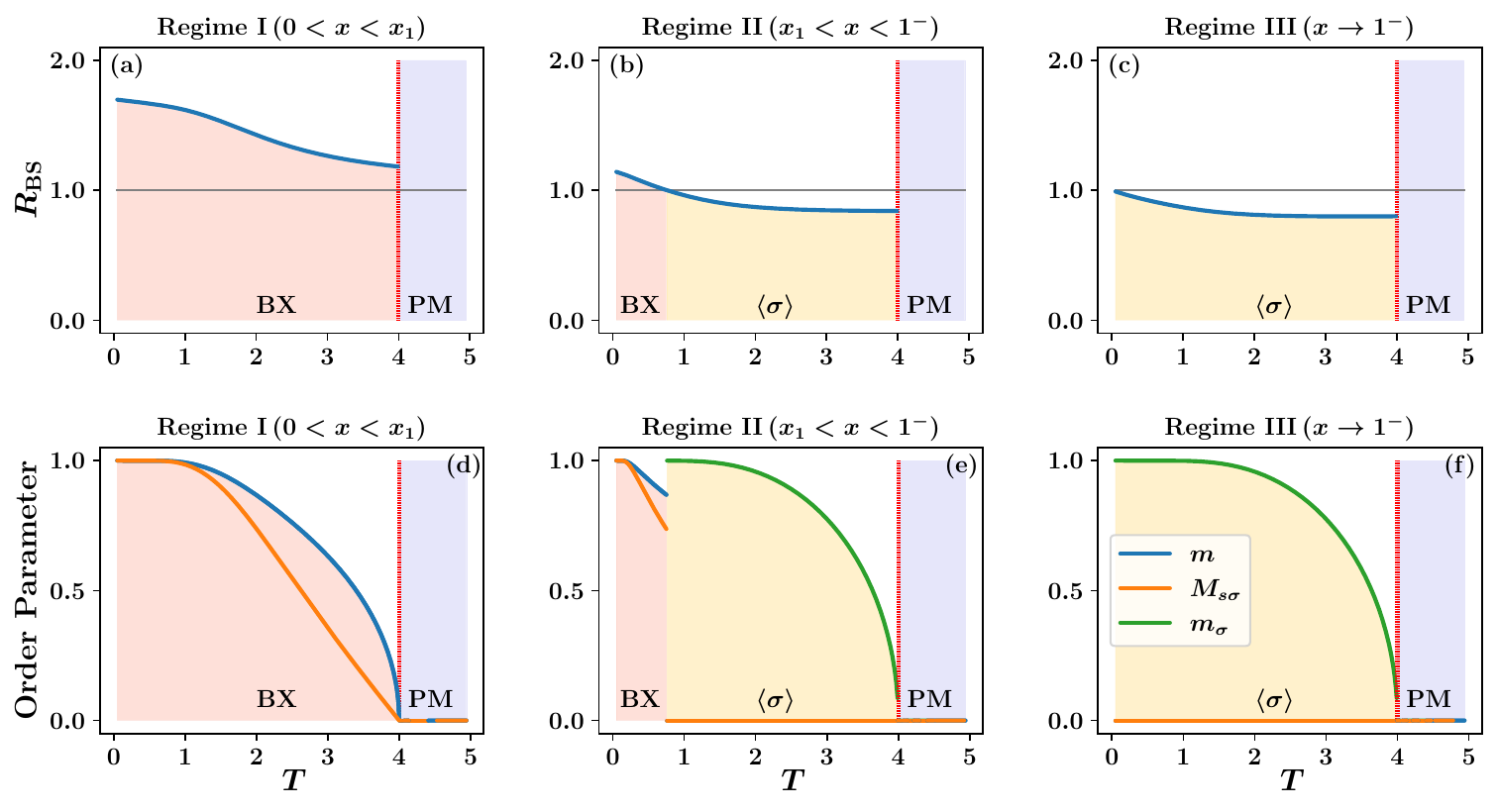}}
    \caption{ (Color online) Temperature dependence of order parameters and free energy ratios in homogeneous networks across different regimes for $0 < x \leq 1$. Panels (a,d) show Regime I ($0 < x < x_1$) with direct Baxter-PM transition. Panels (b,e) display Regime II ($x_1 < x < 1^-$) exhibiting Baxter-$\langle\sigma\rangle$-PM transitions. Panels (c,f) illustrate Regime III ($x \to 1^-$) with $\langle\sigma\rangle$-PM transition. Red dotted lines mark critical temperatures throughout.
    }
    \label{fig:fig10}
\end{figure}

In homogeneous networks, unlike random scale-free networks with $3<\lambda<4$, $R_{\BS}(T)$ in Regime I shows a monotonic decrease with increasing temperature from $T_0$ to $T_c$ ($|t| \to 0$), as illustrated in Fig.~\ref{fig:fig10}.

\clearpage
\newpage

\section{$R_{\BA}(T)$ Behavior}
\label{seca:R_function_strong}

The behavior of $R_{\BA}(T)$ exhibits distinct temperature dependence across different network types, particularly near absolute zero ($T_0$) and critical temperature ($T_c$).

\subsection{For random scale-free networks with $3<\lambda<4$}

In random scale-free networks with $3<\lambda<4$, the behavior of $R_{\BA}(T)$ exhibits non-monotonic temperature dependence, leading to unconventional phase transitions.

\subsubsection{Behavior near $T_0$}
Near $T_0$, the ratio $R_{\BA}(T)$ takes the form:
\begin{align}
   R_{\BA}(T) 
   & = \frac{f_{\Baxter, \min}}{f_{\AF, \min}} = \frac{|f_{\Baxter, \min}|}{|f_{\AF, \min}|}
   \simeq \frac{2 - x}{x + 2 T \log 2 / \kbar} + \text{H.O.}, 
\label{eq:freeEnergyDensity_AT_0_ratio_AF}
\end{align}

\subsubsection{Behavior near $T_c$}
For temperatures above $T_c$, the value of $|f_{\Baxter, \min}|$ vanishes, resulting in $R_{\BA}(T) = 0$. For temperatures slightly below $T_c$, $R_{\BA}(T)$ can be expressed as:
\begin{align}
   R_{\BA}(T)
   & = \frac{f_{\Baxter, \min}}{f_{\AF, \min}} 
   \simeq\frac{(\lambda - 1) C_{35}}{\left|f_{\AF, \min}({T = T_c})\right|} \left(\frac{|t|}{(\lambda-1)C_{35} \kbar / T}\right)^{(\lambda-1)/(\lambda-3)} + \text{H.O.}
   \label{eq:freeEnergyDensity_AT_AF_C_ratio_SF}
\end{align}
where $t = (T - T_c)$. 

The expansion of $R_{\BA}(T)$ considers only the constant term of $|f_{\AF, \min}|$, as $x T_c > T_c$ ensures a finite nonzero value of $f_{\AF, \min}$ at $T_c$. Higher-order $|t|$-dependence of $|f_{\AF, \min}|$ does not contribute to the leading behavior.

\subsubsection{Temperature dependence implications}
Near $x \approx 1$, the value of $R_{\BA}(T)$ exhibits non-monotonic behavior. This non-monotonic behavior induces sequential phase transitions from AF phase to Baxter phase and back from Baxter phase to AF phase, as demonstrated in Fig.~\ref{fig:phase_diagram}(m)--(o).

\subsection{Homogeneous networks}

For homogeneous networks, $R_{\BA}(T)$ demonstrates a simpler, monotonic behavior that contrasts with the random scale-free network case.

\subsubsection{Behavior near $T_0$}
Near $T_0$, the ratio $R_{\BA}(T)$ takes the form:
\begin{align}
   R_{\BA}(T) 
   & = \frac{f_{\Baxter, \min}}{f_{\AF, \min}} = \frac{|f_{\Baxter, \min}|}{|f_{\AF, \min}|}
   \simeq \frac{2 - x}{x + 2 T \log 2 / \kbar} + \text{H.O.}. 
\end{align}

\subsubsection{Behavior near $T_c$}
For temperatures below $T_c$, $|f_{\Baxter, \min}|$ vanishes, yielding $R_{\BA}(T) = 0$. For temperatures slightly below $T_c$, $R_{\BA}(T)$ takes the form:
\begin{align}
   R_{\BA}(T)
   & = \frac{f_{\Baxter \min}}{f_{\AF, \min}} 
   \simeq\frac{2}{3 \left|f_{\AF, \min}({T = T_c})\right|} \frac{1 - 4 x}{1 - x} |t|^{2} + \text{H.O.}
   \label{eq:freeEnergyDensity_AT_AF_C_ratio_LMF}
\end{align}
where $t = (T - T_c)$. 

As in the random scale-free case, the expansion considers only the constant term of $|f_{\AF, \min}|$.

\subsubsection{Temperature dependence implications}
\begin{figure}[ht]
    \resizebox{1.0\columnwidth}{!}{\includegraphics{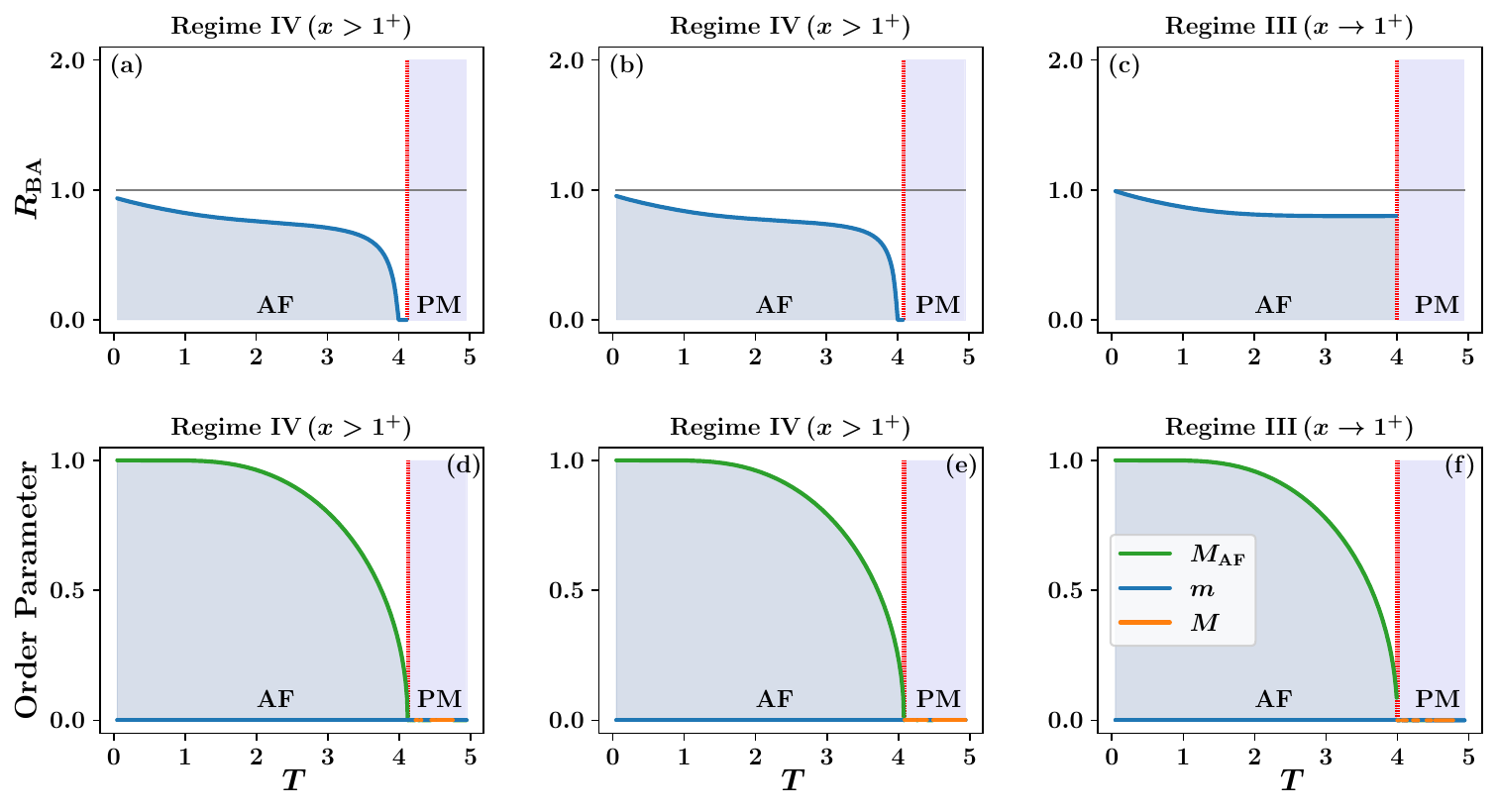}}
    \caption{ (Color online) Temperature dependence of order parameters and free energy ratios in homogeneous networks across different regimes for $x \geq 1$. Panels (a,d) and (b,e) show Regime IV ($x > 1^+$) with direct AF-PM transition. Panels (c,f) illustrate Regime III ($x \to 1^+$) with AF-PM transition. The monotonic decrease of $R_{\BA}(T)$ indicates AF phase dominance over the Baxter phase at all temperatures. Red dotted lines mark critical temperatures.
    }
    \label{fig:fig11}
\end{figure}

In homogeneous networks, unlike random scale-free networks with $3<\lambda<4$, $R_{\BA}(T)$ in $x>1$ shows a monotonic decrease with increasing temperature from $T_0$ to $T_c < T_c^{'}$ ($|t| \to 0$), as illustrated in Fig.~\ref{fig:fig11}.

\clearpage
\newpage

\section{$\lambda$-Dependence of $R_{\BS}(T)$}
\label{seca:lambda-Dependence}

This section examines the free energy density ratios across different $\lambda$ values. We begin by deriving the local minimum of $f$, given in terms of $m^{*} / T$. This step is essential for obtaining Eq.\eqref{eq:freeEnergyDensity_AT_C_ratio_SF_5_7} and Eq.\eqref{eq:freeEnergyDensity_AT_C_ratio_SF_7}, which govern the behavior of $R_{\BS}(T)$ near the transition temperature.
\begin{align}
m^{*} / T = \left(- \frac{3 \kbar |t|}{K_2 \langle k^4 \rangle}\frac{1 + x}{1 + (3\E(\lambda) + 1)x}\right)^{1/2} \text{ for Baxter}, \quad m^{*} / T = \left(- \frac{3 \kbar |t|}{K_2 \langle k^4 \rangle}\right)^{1/2} \text{ for $\sigma$}, \quad \text{ where, \, }{\E (\lambda) = \frac{\langle k^3 \rangle^{2}}{\kkbar\langle k^4 \rangle},}.
\label{eq:order_parameter_AT_C_SF_5}
\end{align}
Notably, these order parameter formulas remain unchanged for $\lambda > 5$, establishing a critical threshold for our analysis. Hence, in what follows, we investigate two additional critical values, $\lambda_{c1}$ and $\lambda_{c2}$.

\subsection{For $\lambda_{c1}$}
Our numerical simulations reveal a significant phenomenon in the range $5 < \lambda < 7$. Within this interval, there exists a specific value $\lambda_{c1}$ where the protrusion of the Baxter phase diminishes to zero at $x=1$.
Based on this observation and utilizing Eq.~\eqref{eq:freeEnergyDensity_AT_SF_5_7}, we can formulate the determining equation for $\lambda_{c1}$:
\begin{align}
R_{\BS}(T) = \frac{|f_{\Baxter, \min}|}{|f_{\langle \sigma \rangle, \min}|} \simeq 2\frac{1+x}{1+(3\E(\lambda)+1)x} = 1,,
\label{eq:freeEnergyDensity_AT_C_ratio_SF_5_7}
\end{align}
This equation is equivalent to Eq.\eqref{eq:freeEnergyDensity_AT_C_ratio_LMF}. Furthermore, by substituting $x=1$, we derive Eq.\eqref{eq:lambda_c1}.

\subsection{For $\lambda_{c2}$}
The second critical value, $\lambda_{c2}$, emerges from our investigation of the temperature dependence of $R_{\BS}(T)$. As discussed in the main text, preventing non-monotonic behavior of $R_{\BS}(T)$ in the Baxter phase requires that the temperature slope of $R_{\BS}(T)$ at $x_{1}$ be negative.
Our numerical analysis indicates that this condition first occurs when $\lambda$ exceeds 7. Using Eq.~\eqref{eq:freeEnergyDensity_AT_SF_7}, we establish the determining equation for $\lambda_{c2}$:
\begin{align}
\frac{d}{d T} R_{\BS}(T) \propto G_{1} - G_{2} - G_{3} \le 0, \quad \because R_{\BS}(T) \simeq 2\frac{1+x}{1+(3\E(\lambda)+1)x} - (G_{1} - G_{2} - G_{3}) |t|,
\label{eq:freeEnergyDensity_AT_C_ratio_SF_7}
\end{align}
where the coefficients $G_i$ are defined as:
\begin{align}
G_{1} &= \frac{2}{45} F_{6} \langle k^6 \rangle, \
G_{2} = -\frac{1}{6}\frac{x \E(\lambda)}{\left(1 + x(1 + 3 \E(\lambda))\right)^2} \frac{\left(\langle k^4 \rangle\right)^2}{\kkbar}, \
G_{3} = \left(\frac{2}{45} + \frac{2}{3}\frac{x}{1+x}\frac{\langle k^3 \rangle \langle k^5 \rangle}{\kkbar \langle k^6 \rangle} \right)\frac{1}{F_{6}^3}\langle k^6 \rangle, \cr
F_{6} &= \left(1 + 3\frac{x}{1+x}\frac{\left(\langle k^3 \rangle\right)^2}{\kkbar \langle k^4 \rangle} \right).
\end{align}

\end{document}